\documentclass[aps,pra,twocolumn,superscriptaddress,longbibliography]{revtex4-2}
\usepackage{graphicx}
\usepackage{amssymb,amsmath}
\usepackage{bm}
\usepackage{dcolumn}
\usepackage{float}
\usepackage[OT1]{fontenc} 
\usepackage{url}
\usepackage{mathrsfs}
\usepackage{slashed,comment}
\usepackage{color}
\usepackage{verbatim}
\usepackage{enumitem}
\usepackage{soul,physics}
\usepackage[driverfallback=dvipdfm]{hyperref}
\hypersetup{pdfpagemode=FullScreen,colorlinks=true,breaklinks,urlcolor=blue,linkcolor=blue,citecolor=blue}

\usepackage{subfigure}
\usepackage{amssymb}
\usepackage{bm}
\usepackage{graphicx}
\usepackage{amsmath,amssymb}
\usepackage{xcolor}

\begin{document}

\title{Few-Magnon Universality beyond Contact Interactions in Long-Range Spin Chains}
  \author{Yiran Huo}
    \affiliation{State Key Laboratory of Surface Physics \& Department of Physics, Fudan University, Shanghai, 200438, China}

  \author{Pengfei Zhang}
  \thanks{PengfeiZhang.physics@gmail.com}
  \affiliation{State Key Laboratory of Surface Physics \& Department of Physics, Fudan University, Shanghai, 200438, China}
  \affiliation{Hefei National Laboratory, Hefei 230088, China}

  \date{\today}

\begin{abstract}
Quantum few-body systems exhibit a rich variety of phenomena that provide fundamental insights into universality. A paradigmatic example is the Efimov effect, in which an infinite tower of three-body bound states emerges with binding energies forming a geometric sequence. Motivated by advances in trapped-ion systems, recent theoretical work has predicted that the Efimov effect can arise in certain long-range quantum spin chains, where magnons exhibit a general dispersion $\epsilon_k=u|k|^z$ with $z\in(0.52,0.88)$ and interact through a contact potential. Nevertheless, whether Efimov universality persists in the presence of experimentally relevant long-range interactions between magnons remains unexplored. In this work, we address this issue by investigating the two- and three-magnon problems with a model interaction potential $V(x)=V_0/(1+|x|^\beta)$. Our results demonstrate that few-body universality persists for any $\beta>1$: In the two-magnon sector, the low-energy expansion of the scattering matrix near a two-body resonance agrees with the universal form predicted by contact-interaction theory. The three-magnon bound-state spectrum exhibits Efimov scaling, with a scaling factor consistent with the theory. We further highlight the experimental relevance of our model, which can be readily realized in trapped-ion systems. Our results pave the way for quantum simulations of universal few-body physics beyond the traditional paradigm of nonrelativistic particles.
\end{abstract}

\maketitle

\section{Introduction}
Quantum few-body problems provide a natural setting for exploring universality, which states that low-energy properties can become insensitive to microscopic details. A prominent example is scattering between nonrelativistic particles interacting through short-range potentials in three dimensions. In the two-body sector, a single parameter, the scattering length \cite{landau1991quantum}, is sufficient to characterize the low-energy physics. When the scattering length diverges, the system is said to be at resonance, and the absence of a characteristic length scale leads to emergent continuous scale invariance. In the three-body sector, this continuous scale invariance can be broken down to discrete scale invariance \cite{Zhai2021Ultracold}, giving rise to the celebrated Efimov effect \cite{EFIMOV1970563}: an infinite tower of three-body bound states with energies satisfying $E_n/E_{n+1}=e^{2\pi/s_0}$. These universal few-body phenomena have attracted broad interest in atomic, nuclear, condensed matter, and particle physics \cite{Hammer:2005bp,Nielsen:2001hbm,Ferlaino2010FortyYO,Braaten:2006vd,Hammer:2010kp,Nishida:2012hf,Naidon:2016dpf,sciadvaau5096,Kievsky:2021ghz,ZHANG2019289}, reflecting their relevance across widely separated energy scales. In particular, connecting the theoretical predictions to cold-atom experiments requires their robustness to power-law interaction tails, since realistic interatomic potentials have van der Waals tails that decay as $1/r^6$~\cite{RevModPhys.82.1225}.

Meanwhile, advances in quantum science and technology have enabled highly controllable simulations of unconventional quantum systems that are otherwise inaccessible. For example, trapped-ion quantum platforms realize quantum spin chains with long-range couplings $J_{mn}\sim J/|m-n|^\alpha$ \cite{wu2023research,2019ApPRv...6b1314B,Foss-Feig:2024blk,2021NatRM...6..892B,10164086,RevModPhys.93.025001,RevModPhys.93.025001,Kim:2010hll,Monroe:2011vee,Schneider2012ExperimentalQS,PhysRevLett.107.150501,Dumitrescu:2021uin,Morong:2021qyr,Zhang:2017kde,Joshi:2021fno,Feng:2022ebi}, whose strength and decay exponent $\alpha\in (0,3)$ can be tuned by controlling the spin-phonon interaction \cite{Monroe2021Programmable, fossfeig2024progresstrappedionquantumsimulation}. Motivated by these developments, several studies have investigated the few-body physics of magnons in long-range quantum spin chains with $U(1)$ spin-rotation symmetry about the $z$ axis~\cite{Sun:2025gew,Sun:2025tdp,Sun:2025kta}. These studies reveal a new paradigm for exploring few-body physics beyond the conventional setting of nonrelativistic particles, as long-range couplings in the $xy$ plane give rise to a more general low-energy dispersion $\epsilon_k=u|k|^z$, with dynamical exponent $z=\min\{2,\alpha-1\}$ for $\alpha>1$. For short-range couplings along the $z$ axis, which give rise to short-range interactions between magnons, Ref.~\cite{Sun:2025gew} predicts continuous scale invariance in the two-body sector at resonance for $z\in(1/2,1)$ and the Efimov effect in the three-body sector for $z\in(0.52,0.88)$. These predictions contrast sharply with conventional scattering theory for nonrelativistic particles in one dimension.

Although short-range couplings between ions have been realized experimentally~\cite{Schuckert:2023pej}, combining them with long-range couplings in the $xy$ plane introduces additional experimental complexity, making the previously studied model favorable in experiments. Indeed, the experimental study of magnon bound states in trapped-ion systems have focused on anisotropic long-range Heisenberg models in which the couplings in all three spin directions share the same power-law decay exponent $\alpha$~\cite{Kranzl2023Magnon}. These experiments motivate a systematic investigation of whether the universal few-body behavior predicted for short-range magnon interactions persists in the presence of long-range interactions. Beyond its experimental relevance, this question is of fundamental theoretical interest, as it probes how the interplay between the dispersion and interaction decay exponents determines the stability of few-body universality and the validity of an effective field theory description based on contact interactions.

\begin{figure}[t]
\centering
\includegraphics[width=0.98\columnwidth]{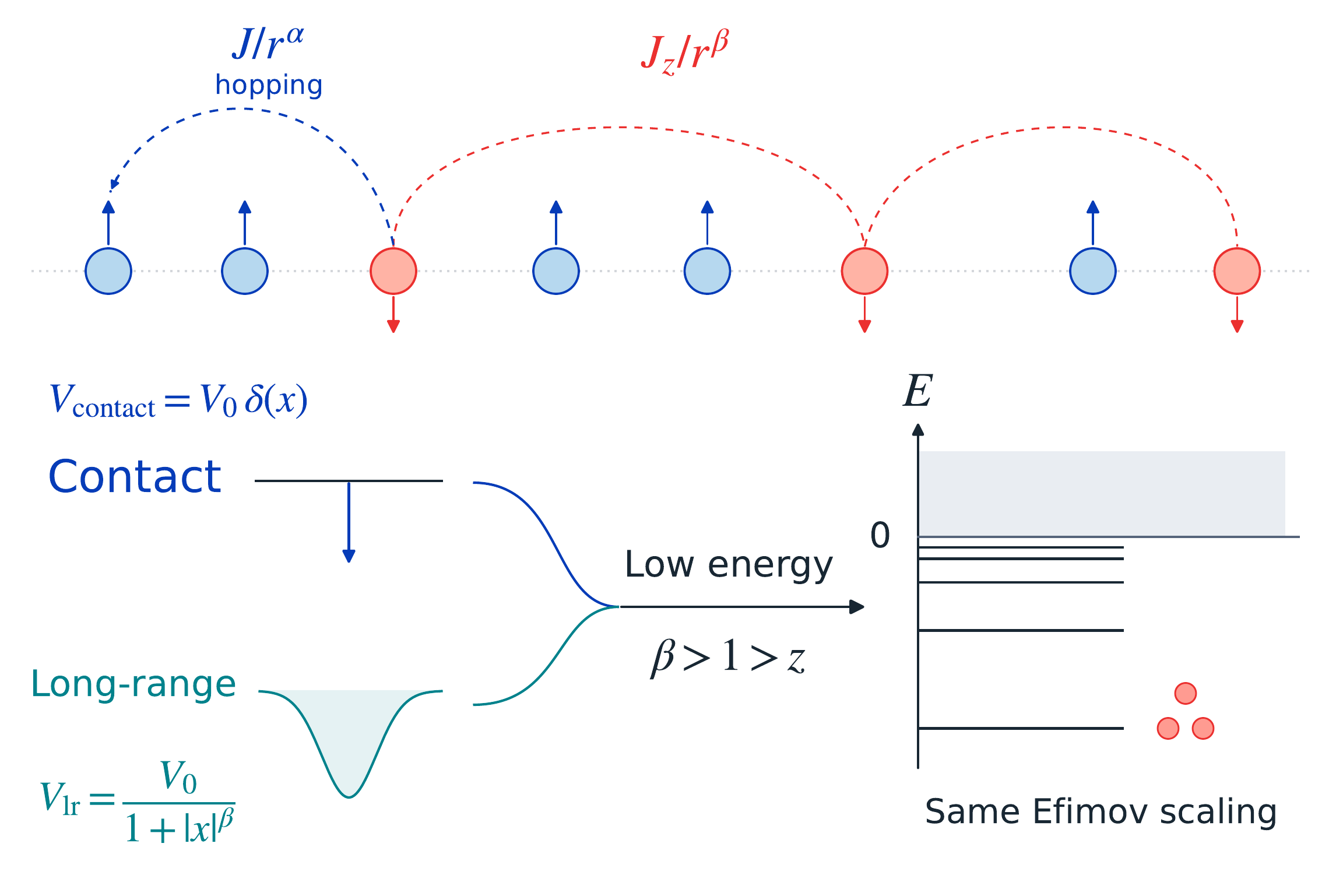}
\caption{Schematic illustration of the long-range quantum spin chain studied in this work. The couplings in the $xy$ plane and along the $z$ axis decay as $r^{-\alpha}$ and $r^{-\beta}$ at large distances, respectively. We focus on $\alpha\in(3/2,2)$, where the dispersion exponent is $z=\alpha-1$ and a nontrivial two-magnon resonance with continuous scale invariance exists. For $\beta>1>z$, the low-energy few-body physics agrees with that of the contact-interaction model, including its predictions for the three-magnon Efimov effect.}
\label{fig}
\end{figure}

In this work, we address this issue by investigating the few-body physics of magnons with a tunable dynamical exponent and long-range interactions described by $V(x)=V_0/(1+|x|^\beta)$, as illustrated in FIG.~\ref{fig}. We focus on the regime $z=\alpha-1\in(1/2,1)$, where a nontrivial two-magnon resonance with continuous scale invariance exists~\cite{Sun2026Efimov}. Combining perturbative analysis with numerical calculations, we show that, for any $\beta>1$, the leading low-energy form of the two-body $T$ matrix agrees with effective theory with contact interactions, and the two-magnon system retains continuous scale invariance at resonance. Numerical calculations in the three-magnon sector further support agreement with the effective theory for the emergence and asymptotic geometric scaling of Efimov bound states. In particular, the experimentally relevant case $\alpha=\beta$ automatically satisfies $\beta>1$ in the regime studied here. These findings clarify how few-body universality survives power-law interaction tails and provide a theoretical basis for exploring universal few-body physics beyond the traditional paradigm of nonrelativistic particles.

\section{Few-Magnon Hamiltonian in Long-Range Spin Chains}
To motivate the magnon problem studied in this work, we consider a class of microscopic models of long-range quantum spin chains. Each site hosts a spin-$1/2$ degree of freedom, described by the Pauli operators $\sigma_n^\alpha$, where $\alpha\in\{x,y,z\}$ and $n\in\mathbb{Z}$. The Hamiltonian reads
\begin{equation}
H=-\sum_{n<m}\left[\frac{J}{|m-n|^\alpha}
\left(\sigma_n^x\sigma_m^x+\sigma_n^y\sigma_m^y\right)
+\frac{J_z}{|m-n|^\beta}\sigma_n^z\sigma_m^z\right],
\label{eq:H}
\end{equation}
where we take $J,J_z>0$. We primarily focus on $\alpha,\beta>1$, which ensures that the energy of generic quantum states is extensive. In the limit $\beta\rightarrow\infty$, the interaction along the $z$ axis is restricted to nearest neighbors, and the model reduces to that studied in Ref.~\cite{Sun:2025gew}. For $\beta=\alpha$, this model becomes the anisotropic long-range Heisenberg model realized in a recent experiment using Floquet engineering~\cite{Kranzl2023Magnon}, up to an overall sign~\footnote{This sign can be compensated by instead studying the high-energy states of the system.}. This implementation can be generalized to $\beta\neq\alpha$ by tuning the decay exponents of the spin-spin interactions. We also present an alternative scheme for realizing $\beta=\alpha$ later in this work.

The Hamiltonian~\eqref{eq:H} is invariant under spin rotations about the $z$ axis, and its eigenstates can be labeled by the total magnetization $M^z=\sum_n \sigma^z_n$. To make contact with conventional scattering problems, we identify the fully polarized spin-up state, $\ket{0} \equiv \bigotimes_n \lvert \uparrow\rangle_n$, as the vacuum and interpret each flipped spin as a particle, referred to as a magnon. The transverse coupling in the first term of the Hamiltonian describes magnon hopping and determines the magnon dispersion. Direct evaluation shows that the energy of the single-magnon state $\ket{k}=\sum_n e^{ikn}\sigma_n^-\ket{0}$, with $\sigma_n^-=\sigma_n^x-i\sigma_n^y$, is given by
\begin{equation}
\epsilon_k
= -\sum_{r=1}^{\infty}\frac{4J}{r^\alpha}\cos(kr)
+\sum_{r=1}^{\infty}\frac{4J_z}{r^\beta}.
\end{equation}
Taking the bottom of the single-particle band as the zero of energy and expanding at small $k$, we obtain the low-energy dispersion~\cite{Sun:2025gew}
\begin{equation}
\epsilon_k\approx u|k|^z,\qquad
z=\min\{\alpha-1,2\}.
\end{equation}
Here, $u>0$ is a constant whose explicit form is not needed in the following discussion. We keep the logarithmic correction that arises at $\alpha=3$ implicit. This dispersion enables the study of bosonic quasiparticles with dynamical exponents $z$ beyond those accessible on traditional experimental platforms. As discussed in the introduction, we focus on the regime $z=\alpha-1\in(1/2,1)$, where a nontrivial two-magnon resonance with continuous scale invariance exists~\cite{Sun2026Efimov}. The coupling along the $z$ axis contributes only a momentum-independent shift to the single-magnon energy. When multiple magnons are excited, however, it generates pairwise interactions of the form $-4J_z/|n-m|^\beta$, which govern their scattering and bound-state formation. Thus, $\alpha$ determines the low-energy dispersion exponent, whereas $\beta$ controls the long-range tail of the two-body interaction.

Combining these ingredients, we describe $N$ magnons in the low-energy continuum limit using the following first-quantized effective Hamiltonian:
\begin{equation}
\hat{H}_{\mathrm{eff}}=\sum_{i=1}^N \hat{T}_i+\sum_{1\leq i<j\leq N}V(\hat{x}_j-\hat{x}_i).
\label{eq:Heff}
\end{equation}
Here, $\hat{x}_i$ is the position operator for the $i$th magnon. $\hat{T}_i=u\left|-i\partial_{x_i}\right|^z$ is the kinetic term, which reproduces the low-energy dispersion $u|k|^z$ in momentum space. The two-body interaction potential $V(x)$ should decay as $1/|x|^\beta$ in the long-distance limit. To guarantee its regularity near $x=0$, we model the interaction in the low-energy regime as
\begin{equation}\label{eq:V}
V(x)=\frac{V_0}{1+|x|^\beta},\qquad V_0<0.
\end{equation}
By tuning $V_0$, we investigate zero-energy resonances at which a two-body bound state reaches the scattering threshold. We then compare the predictions of this effective Hamiltonian for the two- and three-body problems with those of the contact-interaction model $V(x)=-g\delta(x)$ under appropriate renormalization relations~\cite{Sun:2025gew}. The aim is to investigate whether the low-energy scattering behavior recovers the universal results of the contact-interaction model.

\section{Two-Magnon Problem}
We first consider the two-magnon problem with $N=2$. The system is translationally invariant, so the total momentum is conserved. We focus on the sector with fixed total momentum, where the two-body wave function takes the form $\Psi(x_1,x_2)=e^{iQ\frac{x_1+x_2}{2}}\psi(x_1-x_2)$. The Hamiltonian for the relative coordinate, denoted as $\hat{x}=\hat{x}_1-\hat{x}_2$, then reduces to
\begin{equation}
\begin{aligned}
\hat{H}&=u\left|\frac{Q}{2}-i\partial_x\right|^z+u\left|\frac{Q}{2}+i\partial_x\right|^z+V(\hat{x})\equiv \hat{H}_0(Q)+\hat{V}.
\end{aligned}
\end{equation}
We solve the scattering problem using the standard Lippmann-Schwinger equation~\cite{landau1991quantum}, which takes the operator form
\begin{equation}
\hat{T}(E,Q)=\hat{V}+\hat{V}\hat{G}_0(E,Q)\hat{T}(E,Q),
\end{equation}
where $\hat{T}(E,Q)$ is the two-body scattering $T$-matrix, $\hat{G}_0(E,Q)=(E^+-\hat{H}_0(Q))^{-1}$ is the free Green's function, with the shorthand $x^+=x+i0^+$. Since magnons are bosons, exchange symmetry requires the relative wave function to be even under $x\rightarrow -x$, so only even-wave scattering contributes. In this subspace, we restrict the independent momentum variable to $k>0$. Inserting a complete set of momentum states, we obtain the explicit form of the Lippmann-Schwinger equation:
\begin{equation}\label{eq:LS}
\begin{aligned}
T(k,k';&E,Q)=V_+(k,k')
\\&+\int_0^{+\infty}\frac{dq}{2\pi}
\frac{V_+(k,q)T(q,k';E,Q)}
{E^+-u\left|\frac{Q}{2}+q\right|^z-u\left|\frac{Q}{2}-q\right|^z}.
\end{aligned}
\end{equation}
Here, $V_+(k,k')=V(k-k')+V(k+k')$ is the symmetrized interaction kernel, where $V(k)$ is the Fourier transform of $V(x)$. In this section, we focus on the sector with vanishing total momentum $Q=0$, and introduce $T(k,k';E)\equiv T(k,k';E,0)$ for conciseness. Then, the poles of the $T$-matrix at $E<0$ correspond to two-body bound states with binding energies $|E|$. A two-body resonance occurs when $V_0$ is tuned to a critical value at which a bound-state pole reaches the threshold $E=0$, causing the $T$-matrix to become singular as $E\to0$. 

\begin{table}[t]
\centering
\caption{Values of $V_0/u$ at the two-body resonance for the potential $V(x)=V_0/(1+|x|^\beta)$.}
\label{tab:V0}
\vspace{5pt}

\begin{minipage}{0.78\columnwidth}
\begin{ruledtabular}
\begin{tabular}{ccc}
$V_0/u$ & $z=0.65$ & $z=0.75$ \\
\hline

$\beta=1.2$ & $-0.749$ & $-0.433$ \\
$\beta=1.5$ & $-0.866$ & $-0.546$ \\
$\beta=2.0$ & $-0.967$ & $-0.644$ \\
\end{tabular}
\end{ruledtabular}
\end{minipage}
\end{table}

For short-range potentials, the $T$-matrix near a two-body resonance was obtained in Ref.~\cite{Sun:2025gew} using the contact potential $V(x)=-g\delta(x)$ with a momentum cutoff $\Lambda$. For completeness, we reproduce the calculation in Appendix~A. The result is given by
\begin{equation}\label{eq:universal}
    T(k,k';E)=\frac{1}{P^{-1}-C_0u^{-1/z}(-E^+)^{\frac{1-z}{z}}}\equiv T_0(E).
\end{equation}
Here, the coefficient $C_0=2^{-1/z}/[2z\sin(\pi/z)]$ depends only on $z$. The parameter $P$ is the sole low-energy scattering parameter, analogous to the scattering length for nonrelativistic particles in three dimensions. It is related to the coupling constant $g$ and the cutoff $\Lambda$ through the renormalization relation. The two-body resonance occurs at $P=\infty$, where $T_0(E)\propto(-E^+)^{\frac{z-1}{z}}$ diverges as $E\to0$. The absence of an additional dimensionful parameters reflects the emergent continuous scale invariance.

\begin{figure}[t]
  \centering
  \includegraphics[width=0.8\columnwidth]{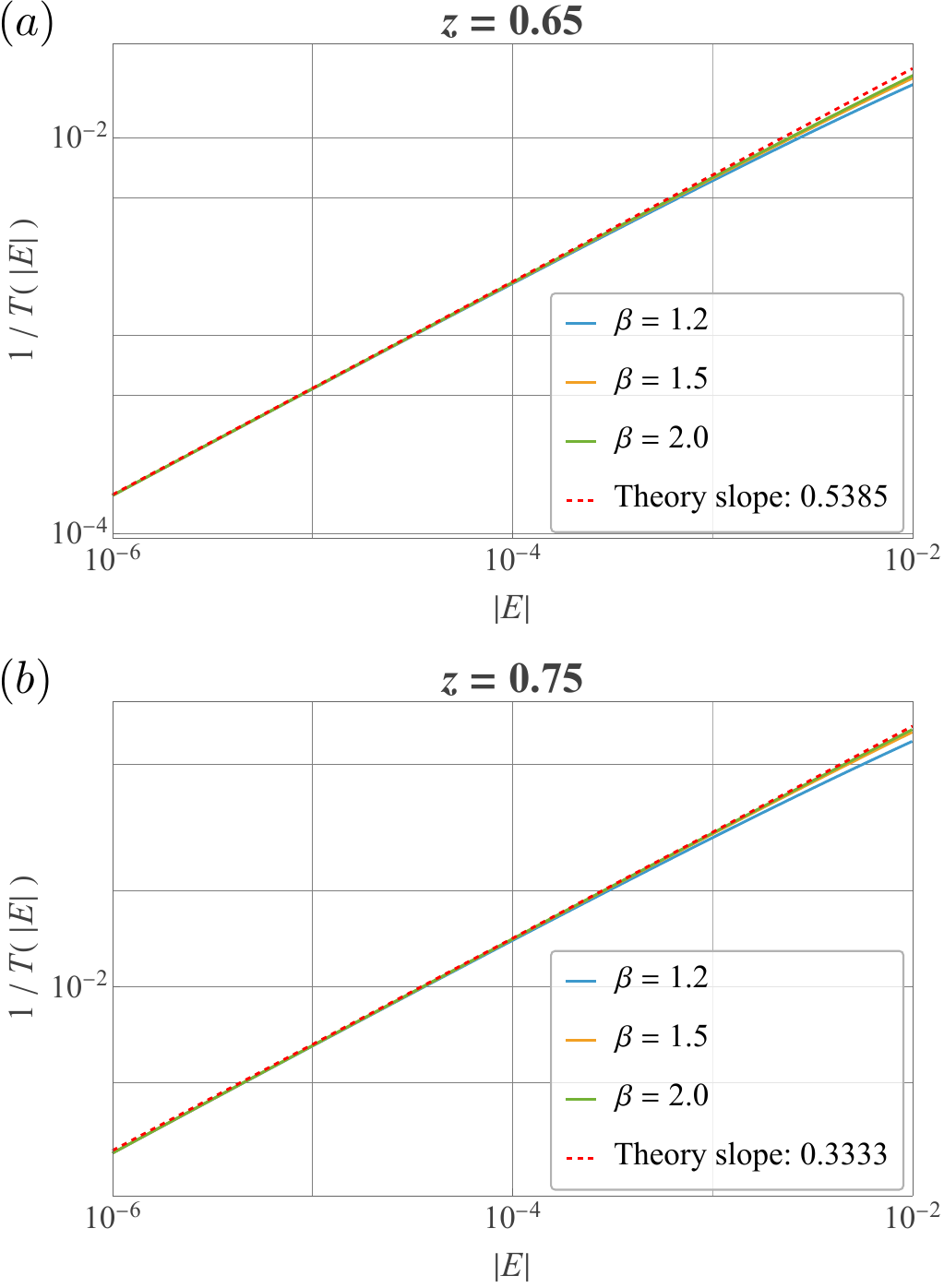}
  \caption{Numerical results for the two-body $T$ matrix at negative energies. The inverse diagonal element $T(k,k;E)^{-1}$, evaluated at $E=-2uk^z$, is plotted against $|E|$ on logarithmic axes for (a) $z=0.65$ and (b) $z=0.75$, with several values of $\beta>1$. We set $u=1$ and use the resonant values of $V_0$ listed in TABLE.~\ref{tab:V0}. The red dashed lines show the contact-interaction predictions, with slope $(1-z)/z$.}
  \label{fig:two-body}
\end{figure}

Next, we return to the long-range potential~\eqref{eq:V}. We perform a perturbative analysis to investigate whether its long-range tail modifies the scaling of the $T$ matrix near the two-body resonance. We separate the potential $V(x)=\frac{V_0}{1+|x|^\beta}$ into a short-range core for $|x|\leq x_0$ and a long-range tail for $|x|>x_0$. With the tail removed, the remaining potential is short-ranged and admits a contact-interaction description at low energies, as in conventional scattering theory~\cite{landau1991quantum}. Upon tuning $V_0$ close to the two-body resonance, the $T$ matrix can be approximated by the universal form~\eqref{eq:universal}. We then restore the long-range tail, denoted by $\delta\hat{V}$, as a perturbation. Varying the Lippmann-Schwinger equation yields the first-order correction to the $T$ matrix:
\begin{equation}
    \delta \hat{T}=(1+\hat{T}_0\hat{G_0})\delta \hat{V}(1+\hat{G_0}\hat{T}_0).
\end{equation}
We focus on the correction to the diagonal matrix element, $\delta T(k,k;E)$. For brevity, we write $\delta T_{kk'}\equiv\delta T(k,k';E)$ and use analogous notation for the other matrix elements, leaving the energy dependence implicit. This leads to
\begin{equation}
\begin{aligned}
\delta &T_{kk}
={}\delta V_{+,kk}
+2\int\frac{dk_1}{2\pi}T_{0,kk_1}G_{0,k_1}\delta V_{+,k_1k}\\
&\ \ \ +\iint\frac{dk_1dk_2}{(2\pi)^2}
T_{0,kk_1}G_{0,k_1}\delta V_{+,k_1k_2}G_{0,k_2}T_{0,k_2k},
\end{aligned}
\label{eq:deltaT}
\end{equation}
where the zeroth-order $T$-matrix element $T_{0,kk'}$ is approximated by Eq.~\eqref{eq:universal}. 

We focus on the on-shell correction $\delta T_{kk}$ at $E=2uk^z$ and perform dimensional analysis at $P=\infty$ to determine its scaling with $k$. At large $|x|$, the interaction tail behaves as $\delta V(x)\sim |x|^{-\beta}$. As in the analysis of the low-energy dispersion, its Fourier transform satisfies $\delta V(q)-\delta V(0)\sim |q|^\gamma$, with $\gamma=\min\{\beta-1,2\}$ for $\beta>1$. The momentum-independent contribution can be absorbed into the short-range interaction, whose strength must be retuned to maintain resonance. In the following infrared power counting, $\delta V$ denotes the remaining momentum-dependent correction. At resonance, the low-energy scaling forms are $G_{0,k}\sim k^{-z}$ and $T_{0,kk}\sim k^{z-1}$. The first term in the perturbative expansion therefore scales as $\delta T^{(1)}_{kk}=\delta V_{kk}\sim k^\gamma$. Applying the same power counting to the infrared contributions of the second and third terms yields the same scaling. Relative to the zeroth-order result, the correction scales as $\delta T_{kk}/T_{0,kk}\sim k^{\gamma+1-z}$ and is therefore subleading when $\gamma>z-1$. Thus, in the regime considered here, $\beta>1>z$, this infrared power counting indicates that the long-range tail does not alter the leading low-energy scaling of two-body scattering.

Numerical results further support this conclusion. For a given energy $E$, we discretize the Lippmann--Schwinger equation into a matrix equation and numerically obtain the $T$-matrix. We determine the values of $V_0/u$ at the two-body resonance by locating the singularity of the solution at $E=0$. The results for several $(\beta,z)$ are listed in Table~\ref{tab:V0}. We then investigate the scaling of $T(k,k;E)$ at these resonant couplings. To avoid the complications associated with a complex $T$-matrix at positive energies, we focus on $E=-2uk^z$, where the $T$-matrix is real. The inverse diagonal element $T(k,k;E)^{-1}$ is plotted against $|E|$ in FIG.~\ref{fig:two-body} for $z=0.65$ and $z=0.75$, with several values of $\beta$ satisfying $\beta>1>z$. This establishes the universality of low-energy two-magnon scattering at resonance for $\beta>1>z$, in agreement with the predictions of contact-interaction theory.

\section{Three-Magnon Problem and Efimov Effect}

We now turn to the three-body problem at the two-body resonance $P=\infty$ and verify that the agreement with contact-interaction theory extends to this sector. For contact interactions, the three-magnon problem has been analyzed using effective field theory, which predicts the Efimov effect for $z\in(0.52,0.88)$~\cite{Sun:2025gew}. However, this formulation does not directly apply to the long-range potential considered here. We therefore start from the first-quantized Hamiltonian~\eqref{eq:Heff} with $N=3$ and derive the one-dimensional Faddeev equations, from which we obtain a generalized Skorniakov--Ter-Martirosian (STM) equation for long-range interactions. For convenience, we recall the Hamiltonian for three identical magnons:
\begin{equation}
\hat{H}_{\mathrm{eff}}=\sum_{i=1}^3 \hat{T}_i+\sum_{1\leq i<j\leq 3}V(\hat{x}_j-\hat{x}_i).
\label{eq:Heff3}
\end{equation}
As in the two-body problem, we focus on the sector with vanishing total momentum. We introduce Jacobi coordinates to distinguish the center-of-mass coordinate from the relative coordinates~\cite{Hammer:2005bp}: $R=\frac{x_1+x_2+x_3}{3}$, $\xi=x_1-x_2$, and $\eta=x_3-\frac{x_1+x_2}{2}$, with conjugate momenta $K=k_1+k_2+k_3$, $q_\xi=\frac{k_1-k_2}{2}$, and $q_\eta=\frac{-k_1-k_2+2k_3}{3}$. Here, $K$ is the total momentum, which we set to zero in the following analysis; $q_\xi$ is the relative momentum of the interacting pair $(1,2)$; and $q_\eta$ is the relative momentum between the pair and the third magnon. In terms of these variables, the kinetic energy $\hat{T}_{3b}=\sum_{i=1}^3 \hat{T}_i$ in the sector with vanishing total momentum is
\begin{equation}
T_{3b}(q_\xi,q_\eta)
=u\left(
\left|\frac{q_\eta}{2}-q_\xi\right|^z
+\left|\frac{q_\eta}{2}+q_\xi\right|^z
+|q_\eta|^z
\right),
\label{eq:T3b}
\end{equation}
where the two relative momenta remain coupled through the nonquadratic dispersion.

We focus on the three-body bound states with energy $E<0$. The Schrödinger equation can be written as
\begin{equation}
|\Phi\rangle
=\hat{G}_{0,3b}(E)(\hat{V}_{12}+\hat{V}_{23}+\hat{V}_{31})|\Phi\rangle,
\end{equation}
where $\hat{G}_{0,3b}(E)=(E-\hat{T}_{3b})^{-1}$ is the free three-body Green function and we introduce the notation $\hat{V}_{ij}=V(\hat{x_i}-\hat{x_j})$ for simplicity. Three Faddev components are defined by separating out the contribution from each term as $|\Phi\rangle = |\psi_{12}\rangle+|\psi_{23}\rangle+|\psi_{31}\rangle$, with
\begin{equation}
|\psi_{ij}\rangle=\hat{G}_{0,\text{3b}}(E)\hat{V}_{ij}|\Phi\rangle,\quad ij=12,23,31.
\end{equation}
Equivalently, we can expand the state $|\Phi\rangle$ on the right-hand side in terms of Faddeev components and use the two-body Lippmann-Schwinger equation to express $|\psi_{12}\rangle$ in terms of the other two components. This leads to
\begin{equation}
|\psi_{12}\rangle
=\frac{1}{2}\hat{G}_{0,3b}(E)\hat{T}_{12}(E-u|q_\eta|^z,-q_\eta)
\bigl(|\psi_{23}\rangle+|\psi_{31}\rangle\bigr).
\end{equation}
Here, $\hat{T}_{12}(E,Q)$ is the two-body scattering $T$ matrix for magnons $(1,2)$, determined by the Lippmann-Schwinger equation~\eqref{eq:LS}. It acts only on their relative-momentum degree of freedom $q_\xi$. The subscript indicates the pair on which the operator acts. The factor of $1/2$ arises due to the symmetrization in Eq.~\eqref{eq:LS}. For identical bosons, the three components are related by particle permutations. Choosing $|\psi\rangle\equiv|\psi_{12}\rangle$ and defining permutation operators $\mathcal{P}_\pm$ through $|\psi_{23}\rangle=\hat{\mathcal{P}}_+|\psi\rangle$ and $|\psi_{31}\rangle=\hat{\mathcal{P}}_-|\psi\rangle$, we reduce the coupled Faddeev equations to
\begin{equation}
|\psi\rangle
=\frac{1}{2}\hat{G}_{0,3b}(E)
\hat{T}_{12}\left(E-u|\hat q_\eta|^z,-\hat q_\eta\right)
(\hat{\mathcal{P}}_++\hat{\mathcal{P}}_-)|\psi\rangle.
\label{eq:faddeev3}
\end{equation}

Finally, to obtain a more compact integral equation, we factor the external free propagator out of the Faddeev component by defining the vertex function $\phi$ through
\begin{equation}
\psi(q_\xi,q_\eta)
=\frac{\phi(q_\xi,q_\eta)}
{E-T_{3b}(q_\xi,q_\eta)}.
\end{equation}
In addition, the Jacobi momenta in different channels are related by transformations whose explicit forms are given in Appendix~B. Combining these ingredients and projecting the Faddeev equation onto momentum space using $\langle q_\xi,q_\eta|$, we obtain the following STM equation for the bound state:
\begin{equation}\label{eq:STM}
\begin{aligned}
\phi(q_\xi,q_\eta)
={}&\int_{-\infty}^{\infty}\frac{dq_\eta'}{2\pi}T\left(
q_\xi,q_\eta'+\frac{q_\eta}{2};
E-u|q_\eta|^z,-q_\eta
\right)
\\
&\times K(q_\eta,q_\eta';E)~\phi\Big(
-q_\eta-\frac{q_\eta'}{2},q_\eta'
\Big),
\end{aligned}
\end{equation}
where 
$$
K(q_\eta,q_\eta';E)
=\frac{1}{
E-u|q_\eta|^z-u|q_\eta'|^z-u|q_\eta+q_\eta'|^z
}
$$
describes the free propagation of the three magnons in the intermediate state. Their momenta are $q_\eta$, $q_\eta'$, and $-q_\eta-q_\eta'$, respectively, ensuring vanishing total momentum. For the long-range potential, the two-body $T$-matrix depends on the relative momenta, so the vertex function generally retains its dependence on both $q_\xi$ and $q_\eta$. The three-body bound-state energies are the values of $E<0$ for which this homogeneous integral equation admits nontrivial solutions.

\begin{figure}[t]
    \centering
    \includegraphics[width=0.99\columnwidth]{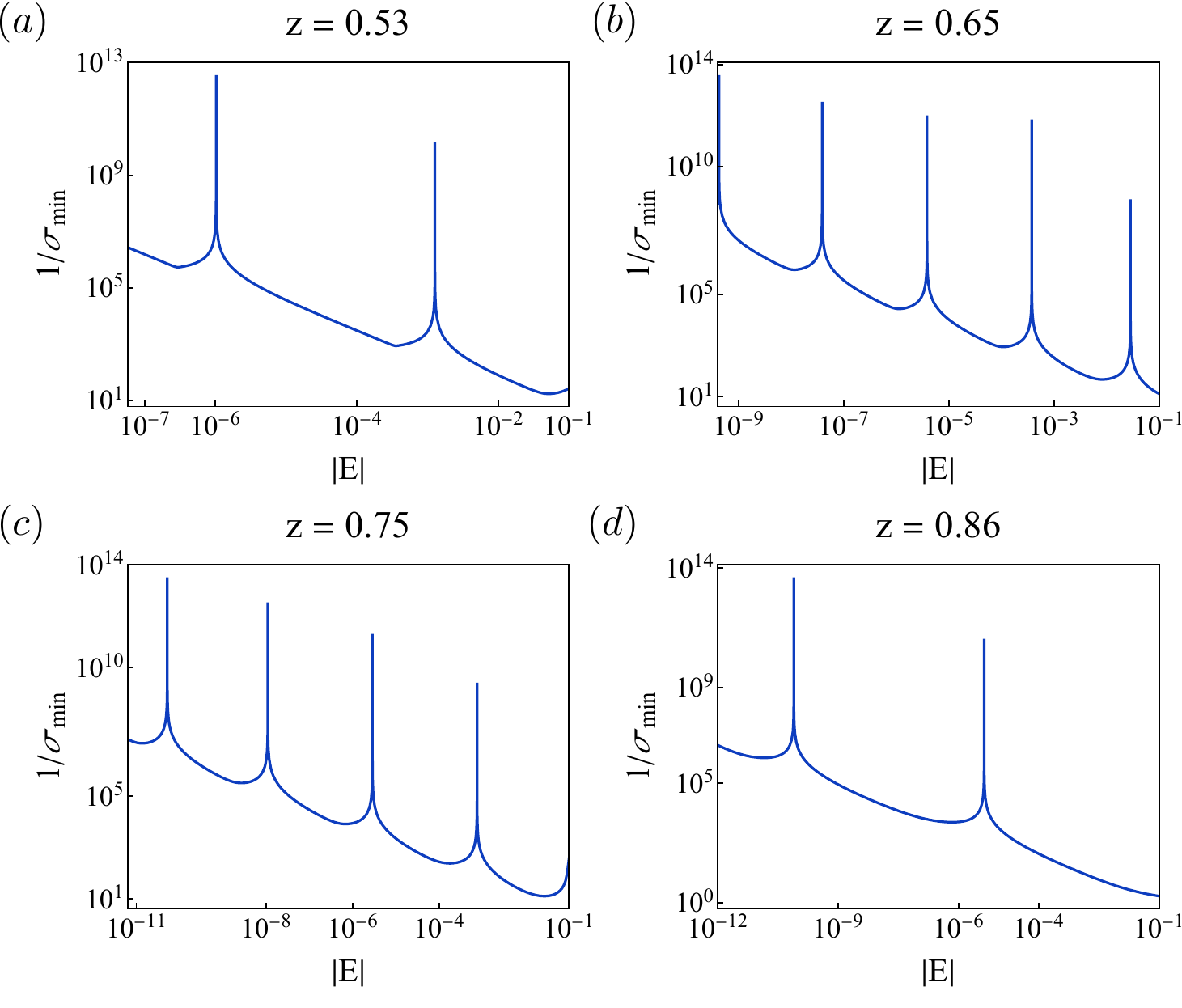}
    \caption{Inverse of the smallest singular value, $1/\sigma_{\min}$, of the matrix $M$ as a function of $|E|$ for (a) $z=0.53$, (b) $z=0.65$, (c) $z=0.75$, and (d) $z=0.86$. Peaks in $1/\sigma_{\min}$ identify the three-body bound-state energies. Equally spaced peaks on a logarithmic energy axis indicate a geometric sequence of bound-state energies.}
    \label{fig:spec}
\end{figure}

We then numerically solve the STM equation~\eqref{eq:STM} to identify three-body bound states. Specifically, we discretize the momentum integral, reducing the equation to the matrix equation $\phi_i=\sum_j A_{ij}(E)\phi_j$. We set $u=1$, tune the interaction strength $V_0$ to the two-body resonance, and scan the total energy $E<0$. A nontrivial solution exists when the smallest singular value of $M(E)=I-A(E)$ vanishes, identifying a three-body bound-state energy. FIG.~\ref{fig:spec} shows the inverse of the smallest singular value, $1/\sigma_{\min}$, as a function of $|E|$ for $z=0.53$, $0.65$, $0.75$, and $0.86$. The peaks in $1/\sigma_{\min}$ locate the three-body bound-state energies. Their approximately equal spacing on a logarithmic energy axis indicates that the binding energies of the resolved bound states form a geometric sequence, which can be parametrized as $E_n=E_0\exp[-z\pi n/s_0(z)]$. These results provide numerical evidence for the Efimov effect in a one-dimensional quantum spin chain with long-range magnon-magnon interactions.

\begin{figure}[htbp]
    \centering
    \includegraphics[width=0.99\columnwidth]{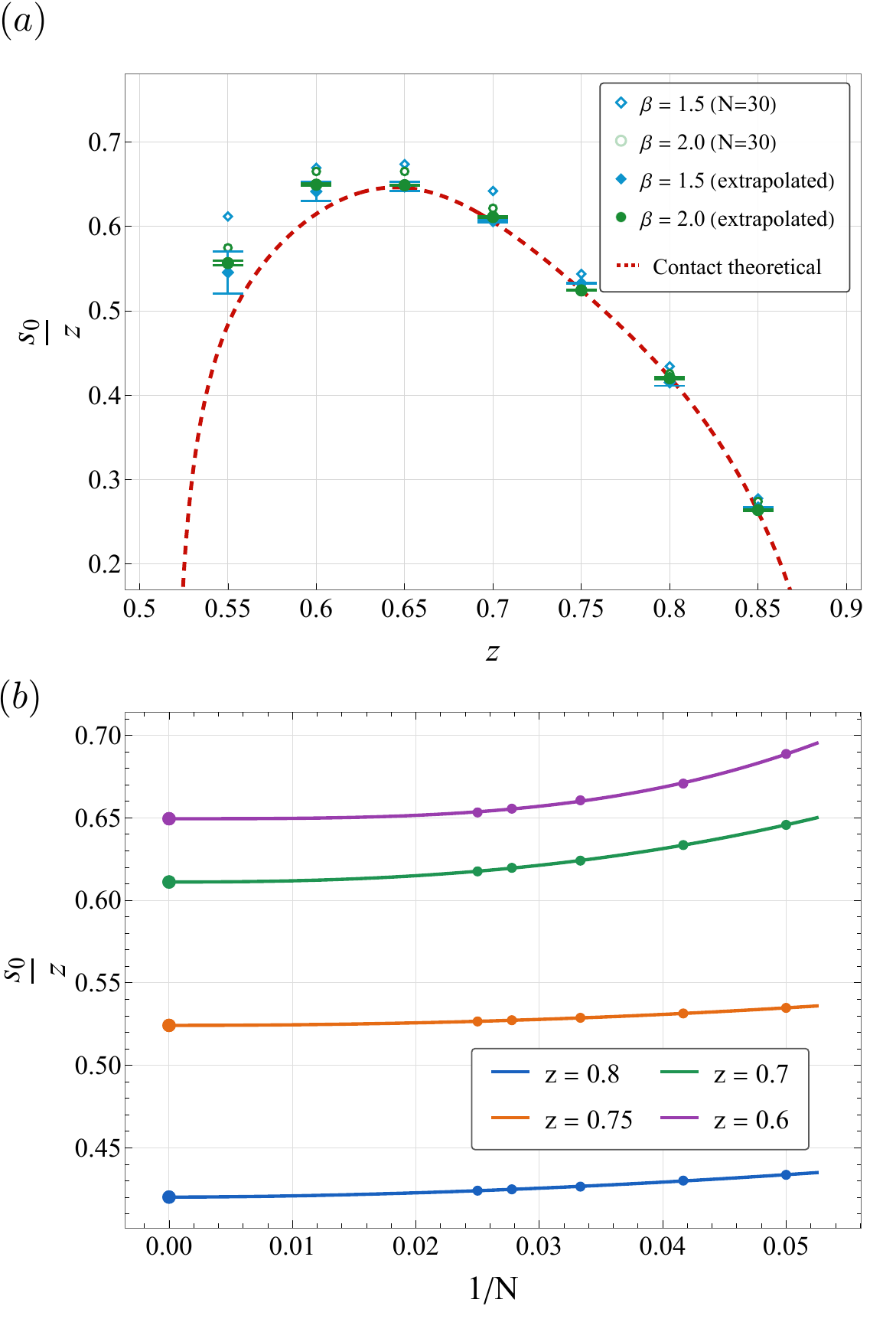}
    \caption{Numerical analysis of the Efimov scaling parameter $s_0(z)/z$ at the two-body resonance. (a) $s_0(z)/z$ as a function of $z$ for long-range magnon-magnon interactions with $\beta=1.5$ and $2.0$. The open symbols show the numerical results obtained on a discrete momentum-space grid with $N=30$, while the filled symbols denote the values extrapolated to $N\to\infty$. The error bars on the extrapolated values represent the uncertainties from the fitting. The red dashed line is the effective field theory prediction for contact interactions. (b) Finite-size extrapolation of $s_0(z)/z$ as a function of $1/N$ for representative values of $z$. Numerical results for $N\in\{20,24,30,36,40\}$ are fitted using $s_0(z)/z=a+b/N^c$, where $a$, $b$, and $c$ are fitting parameters.}
    \label{fig:efimov}
\end{figure}

We further compare the extracted scaling parameter $s_0(z)/z$ with the prediction of contact-interaction theory. As shown in Ref.~\cite{Sun:2025gew}, $s_0(z)$ for short-range potentials is determined by solving the equation
\begin{equation}
\begin{aligned}
&\int_{-\infty}^{\infty}\frac{dx}{2\pi}
\frac{|x|^{z/2-1+i s_0(z)}}
{1+|x|^z+|1-x|^z}=B(z),
\end{aligned}
\label{eq:s0}
\end{equation}
with constant $B(z)=\int_0^\infty\frac{dx}{2\pi}
\Big[
\frac{1}{2x^z}
-\frac{1}{
1+\left(\frac12+x\right)^z
+\left|\frac12-x\right|^z}
\Big]$. FIG.~\ref{fig:efimov}(a) presents the numerical and extrapolation results for $s_0(z)/z$ for long-range interaction potentials with $\beta=1.5$ and $2.0$. Specifically, to resolve the characteristic oscillatory behavior of the low-energy three-magnon wave function \cite{Chen2025Exact}, we introduce the variable $\tau=\text{arcsinh}(k/k^*)$, where, for each energy $E$ in the low-energy scan, the characteristic momentum scale $k^*$ is determined by $-E=u(k^*)^z$. We then discretize $\tau$ on a uniform grid, which corresponds to a nonlinear discretization of momentum space. An inevitable consequence of this discretization scheme is that shallower bound states are resolved by an effectively coarser momentum grid. To minimize the associated discretization effects, we therefore take the value of $s_0/z$ extracted from the spacing between the two deepest-energy peaks in FIG.~\ref{fig:spec} as the representative numerical result. The results obtained with $N=30$ momentum-grid points are shown as open symbols in FIG.~\ref{fig:efimov}(a).
The numerical results agree reasonably well with the effective field theory predictions for contact interactions, but remain systematically higher at relatively small $z$.

We attribute the discrepancy to finite discretization effects. To test this, we calculate $s_0(z)/z$ for each value of $z$ in the scan using $N\in\{20,24,30,36,40\}$ momentum-grid points. The resulting $N$-dependence is fitted by $s_0(z)/z=a+b/N^c$, where $a$, $b$, and $c$ are fitting parameters, as shown in FIG.~\ref{fig:efimov}(b). The intercept $a$ gives the extrapolated value of $s_0(z)/z$ in the limit of $N\to\infty$. The extrapolated results are also shown in FIG.~\ref{fig:efimov}(a) with the filled symbols, and exhibit a clear shift toward the effective field theory prediction. The error bars on the extrapolated values represent the uncertainties associated with the fitting. Therefore, within the parameter regime studied here, where $\beta>1$ and the Efimov effect occurs, these results therefore support the robustness of three-magnon universality to power-law tails in the interaction potential.

\section{Experimental Realization}
Our results show that universal few-body phenomena, including emergent continuous scale invariance in the two-magnon sector at resonance and the magnon Efimov effect, can arise in the anisotropic long-range Heisenberg model with Hamiltonian
\begin{equation}
H=\sum_{n<m}\frac{J}{|m-n|^\alpha}
\left(\sigma_n^x\sigma_m^x+\sigma_n^y\sigma_m^y+\Delta \sigma_n^z\sigma_m^z\right).
\label{eq:Hf}
\end{equation}
The high-energy states of this model correspond to the low-energy states of the Hamiltonian~\eqref{eq:H} with $\beta=\alpha$ and anisotropy parameter $\Delta=J_z/J$. This model has been realized using Floquet engineering in Ref.~\cite{Kranzl2023Magnon}. Here, we present an alternative implementation in trapped-ion systems that avoids the need for multiple pulses while retaining a tunable anisotropy $\Delta$ by generalizing the approach for realizing the long-range XY model \cite{Feng:2022ebi}. 

Specifically, we start from a long-range Ising interaction generated by bichromatic Raman fields through virtual phonon exchange and supplement it with uniform transverse and longitudinal fields~\cite{Monroe2021Programmable,Feng:2022ebi}. The resulting Hamiltonian is
\begin{equation}
H_{d}=
\sum_{n<m}\frac{K}{|m-n|^\alpha}\sigma_n^x\sigma_m^x
+\sum_n\left(\delta\sigma_n^z+\Omega\sigma_n^x\right),
\label{eq:driven_Ising}
\end{equation}
where $K>0$ denotes the Ising coupling strength. The transverse field $\delta$ is controlled by a common frequency offset of the bichromatic tones, while the longitudinal field $\Omega$ is generated by an additional carrier drive whose phase is aligned with the Ising axis and whose frequency matches the center frequency of the bichromatic pair. These fields can be applied simultaneously, allowing the strength and orientation of the total effective field to be controlled continuously.

To obtain the desired interaction, we align the spin quantization axis with the effective field by introducing $\tau_n^z=(\delta\sigma_n^z+\Omega\sigma_n^x)/B$, $\tau_n^x=(\delta\sigma_n^x-\Omega\sigma_n^z)/B$, and $\tau_n^y=\sigma_n^y$, where $B=\sqrt{\delta^2+\Omega^2}$. When $B$ greatly exceeds the interaction scale $\max_n\sum_{m\neq n}K/|m-n|^\alpha$, while remaining small compared with the relevant phonon detunings, we can apply the secular approximation in the interaction picture generated by $B\sum_n\tau_n^z$. Discarding the rapidly oscillating terms yields
\begin{equation}
H_{\mathrm{sec}}=
\sum_{n<m}\frac{K}{|m-n|^\alpha}
\left[
\frac{\delta^2}{2B^2}
\left(\tau_n^x\tau_m^x+\tau_n^y\tau_m^y\right)
+\frac{\Omega^2}{B^2}\tau_n^z\tau_m^z
\right].
\label{eq:secular_XXZ}
\end{equation}
After relabeling the rotated spin axes, this reproduces the anisotropic Heisenberg Hamiltonian \eqref{eq:Hf}, with $J=\frac{K\delta^2}{2(\delta^2+\Omega^2)}$ and $\Delta=\frac{2\Omega^2}{\delta^2}$. The limit $\Omega=0$ recovers the long-range XY model, whereas a finite longitudinal field produces a tunable positive anisotropy. Thus, varying $\Omega/\delta$ allows access to the desired XXZ model while preserving the spatial decay of the interactions and requiring only continuous driving.

\section{Discussions}
In this work, we investigate the robustness of universal two- and three-magnon physics to long-range interactions, extending previous studies of contact interactions in long-range spin chains~\cite{Sun:2025gew}. We consider an effective model with dispersion $\epsilon_k=u|k|^z$ and interaction potential $V(x)=V_0/(1+|x|^\beta)$. Using perturbative analysis and numerical solutions of the Lippmann--Schwinger equation, we show that, within the regime $1/2<z<1$ studied here, corrections from the interaction tail are subleading to the contact-interaction contribution near a two-body resonance for $\beta>1$. Consequently, the two-magnon sector retains its universal low-energy behavior and emergent continuous scale invariance at resonance.

We further investigate the three-magnon problem using the Faddeev decomposition and derive an STM-type integral equation involving the two-body $T$ matrix. Numerical solutions at the two-body resonance reveal a geometric sequence of bound-state energies within the range of dispersion exponents supporting the Efimov effect. Although residual numerical deviations remain, convergence tests support agreement with the scaling factors predicted by effective field theory with contact interactions. Together, our two- and three-body results provide evidence that few-magnon universality survives power-law interaction tails for $\beta>1$, paving the way for experimental studies of universal few-body physics on state-of-the-art quantum simulation platforms.

We conclude by outlining several directions for future research. First, investigating decoherence and loss in long-range spin chains could clarify how dissipation modifies resonant scattering and the Efimov spectrum. Such studies would help assess the experimental observability of these phenomena and establish when a complex scattering parameter provides an effective description~\cite{Hammer:2005bp}. Second, extending our analysis to four or more magnons could reveal whether additional universal bound states accompany the Efimov trimers and how their properties depend on the dynamical exponent. Finally, exploring the algebraic structure underlying scale invariance at resonance for general dynamical exponents $z$ could reveal universal constraints on magnon dynamics.

\vspace{5pt}
  \textit{Acknowledgement.}
  We thank Ning Sun for helpful discussions. This project is supported by the Quantum Science and Technology-National Science and Technology Major Project 2024ZD0300101, the Shanghai Rising-Star Program under grant number 24QA2700300, the NSFC under grant 12374477, the Xuemin Institute of Advanced Studies at Fudan University, and the Shanghai Qiguang Natural Science Development Foundation.

\appendix

\section{Two-body $T$ matrix for contact interactions}
\label{app:T}

For completeness, we derive the universal two-body $T$ matrix for the contact potential $V(x)=-g\delta(x)$, with $g>0$. We use the even-parity convention of Eq.~\eqref{eq:LS}, in which the independent relative momentum is positive and the symmetrized interaction kernel is $V_+(k,k')=-2g$. The $T$ matrix is therefore independent of the incoming and outgoing relative momenta, and we write $T(k,k';E,Q)=T_0(E,Q)$. Introducing a momentum cutoff $\Lambda$, the Lippmann--Schwinger equation reduces to
\begin{equation}
T_0(E,Q)=-2g-2g\mathcal{I}_\Lambda(E,Q)T_0(E,Q),
\label{eq:app_contact_LS}
\end{equation}
where
\begin{equation}
\mathcal{I}_\Lambda(E,Q)
=\int_0^\Lambda\frac{dq}{2\pi}
\frac{1}{E^+-u|Q/2+q|^z-u|Q/2-q|^z}.
\label{eq:app_loop_Q}
\end{equation}
Here, $E^+=E+i0^+$. It follows that
\begin{equation}
T_0(E,Q)^{-1}=-\frac{1}{2g}-\mathcal{I}_\Lambda(E,Q).
\label{eq:app_contact_inverse}
\end{equation}

We first focus on vanishing total momentum and introduce $T_0(E)\equiv T_0(E,0)$ and $\mathcal{I}_\Lambda(E)\equiv\mathcal{I}_\Lambda(E,0)$. For $E<0$, the integral has no singularities on the real momentum axis. Its energy-independent part is
\begin{equation}
\mathcal{I}_\Lambda(0)
=-\int_0^\Lambda\frac{dq}{2\pi}\frac{1}{2uq^z}
=-\frac{\Lambda^{1-z}}{4\pi u(1-z)}.
\label{eq:app_loop_zero}
\end{equation}
Subtracting this contribution gives
\begin{equation}
\begin{aligned}
\mathcal{I}_\Lambda(E)-\mathcal{I}_\Lambda(0)
&=\int_0^\Lambda\frac{dq}{2\pi}
\left(\frac{1}{E-2uq^z}+\frac{1}{2uq^z}\right)\\
&=\frac{(2u)^{-1/z}}{2\pi z}|E|^{\frac{1-z}{z}}
\int_0^{2u\Lambda^z/|E|}
\frac{x^{1/z-2}}{1+x}\,dx,
\end{aligned}
\label{eq:app_subtracted_loop}
\end{equation}
where we have introduced $x=2uq^z/|E|$. For $1/2<z<1$, the remaining integral converges at both limits when its upper limit is extended to infinity. Using
\begin{equation}
\int_0^\infty\frac{x^{a-1}}{1+x}\,dx
=\frac{\pi}{\sin(\pi a)},\qquad 0<a<1,
\end{equation}
with $a=1/z-1$, we obtain
\begin{equation}
\begin{aligned}
\mathcal{I}_\Lambda(E)
={}&-\frac{\Lambda^{1-z}}{4\pi u(1-z)}
+C_0u^{-1/z}|E|^{\frac{1-z}{z}}\\
&+O\!\left(\frac{|E|}{u^2\Lambda^{2z-1}}\right),
\end{aligned}
\label{eq:app_loop_expansion}
\end{equation}
where
\begin{equation}
C_0=-\frac{2^{-1/z}}{2z\sin(\pi/z)}>0.
\label{eq:app_C0}
\end{equation}
The cutoff-dependent correction is subleading in the low-energy limit $|E|\ll 2u\Lambda^z$.

We absorb the energy-independent cutoff dependence into the scattering parameter $P$ through the renormalization relation
\begin{equation}
P^{-1}=-\frac{1}{2g}
+\frac{\Lambda^{1-z}}{4\pi u(1-z)}.
\label{eq:app_renormalization}
\end{equation}
Substituting into Eq.~\eqref{eq:app_contact_inverse} and analytically continuing to general $E$ gives, to leading order at low energies,
\begin{equation}
T_0(E)=\frac{1}{P^{-1}-C_0u^{-1/z}(-E^+)^{\frac{1-z}{z}}},
\label{eq:app_universal_T}
\end{equation}
which reproduces the universal form used in the main text. The fractional power is defined by continuation from $E<0$ with the retarded prescription $E^+=E+i0^+$. At the two-body resonance, $P^{-1}=0$, this expression reduces to
\begin{equation}
T_0(E)=-\frac{u^{1/z}}{C_0}(-E^+)^{\frac{z-1}{z}},
\end{equation}
exhibiting the scale-invariant energy dependence discussed in the main text.

The same renormalization relation applies at finite total momentum $Q$. Taking $\Lambda\to\infty$ at fixed $P$ gives the convergent expression
\begin{equation}
\begin{aligned}
T_0&(E,Q)^{-1}
={}P^{-1}-\int_0^\infty\frac{dq}{2\pi}\\
&\times\left[
\frac{1}{E^+-u|\frac{Q}{2}+q|^z-u|\frac{Q}{2}-q|^z}
+\frac{1}{2uq^z}
\right].
\end{aligned}
\label{eq:app_finite_Q}
\end{equation}
This form retains the total-momentum dependence required when the two-body $T$ matrix is embedded in the three-magnon problem. In particular, a spectator with momentum $q_\eta$ leaves the interacting pair with energy $E-u|q_\eta|^z$ and total momentum $-q_\eta$.

\section{Jacobi momentum transformations}
\label{app:Jacobi}

In this appendix, we derive the transformations between the Jacobi momenta of the three Faddeev channels and explain how they enter Eq.~\eqref{eq:STM}. We work in the sector with vanishing total momentum, $k_1+k_2+k_3=0$. When magnon $i$ is the spectator and magnons $(j,k)$ form the interacting pair, we define
\begin{equation}
q_{\xi,i}=\frac{k_j-k_k}{2},\qquad
q_{\eta,i}=\frac{2k_i-k_j-k_k}{3}=k_i,
\end{equation}
where $(i,j,k)$ runs over the cyclic permutations of $(3,1,2)$. The variables used in the main text correspond to spectator 3: $q_\xi=q_{\xi,3}$ and $q_\eta=q_{\eta,3}$. The individual momenta are therefore
\begin{equation}
k_1=q_\xi-\frac{q_\eta}{2},\qquad
k_2=-q_\xi-\frac{q_\eta}{2},\qquad
k_3=q_\eta.
\label{eq:app_individual_momenta}
\end{equation}

For the channel with spectator 1, substitution into the definitions above gives
\begin{equation}
q_{\xi,1}=-\frac{q_\xi}{2}-\frac{3q_\eta}{4},\qquad
q_{\eta,1}=q_\xi-\frac{q_\eta}{2}.
\end{equation}
Similarly, the channel with spectator 2 has
\begin{equation}
q_{\xi,2}=-\frac{q_\xi}{2}+\frac{3q_\eta}{4},\qquad
q_{\eta,2}=-q_\xi-\frac{q_\eta}{2}.
\end{equation}
These transformations can be written in matrix form as
\begin{equation}
\begin{pmatrix}
q_{\xi,1}\\q_{\eta,1}
\end{pmatrix}
=
\begin{pmatrix}
-1/2&-3/4\\1&-1/2
\end{pmatrix}
\begin{pmatrix}
q_\xi\\q_\eta
\end{pmatrix},
\label{eq:app_Jacobi_1}
\end{equation}
and
\begin{equation}
\begin{pmatrix}
q_{\xi,2}\\q_{\eta,2}
\end{pmatrix}
=
\begin{pmatrix}
-1/2&3/4\\-1&-1/2
\end{pmatrix}
\begin{pmatrix}
q_\xi\\q_\eta
\end{pmatrix}.
\label{eq:app_Jacobi_2}
\end{equation}
Both matrices have determinant one, so the integration measure is unchanged. These relations follow directly from the definitions of the Jacobi momenta and do not require a quadratic dispersion.

To connect these transformations to the STM equation, let $p>0$ denote the intermediate relative momentum in the pair $(1,2)$. The two permutation terms involve the vertex functions
\begin{equation}
\begin{aligned}
&\phi\left(-\frac{p}{2}-\frac{3q_\eta}{4},p-\frac{q_\eta}{2}\right),\\
&\phi\left(-\frac{p}{2}+\frac{3q_\eta}{4},-p-\frac{q_\eta}{2}\right),
\end{aligned}
\end{equation}
respectively. In the first term, we set $q_\eta'=p-q_\eta/2$, which covers $q_\eta'>-q_\eta/2$. In the second term, we set $q_\eta'=-p-q_\eta/2$, which covers $q_\eta'<-q_\eta/2$. Bose symmetry requires $\phi(-q_\xi,q_\eta)=\phi(q_\xi,q_\eta)$, so both terms can be written using the same vertex function,
\begin{equation}
\phi\left(-q_\eta-\frac{q_\eta'}{2},q_\eta'\right).
\end{equation}
We similarly extend the even-parity $T$ matrix to signed relative momenta by defining $T(k,k';E,Q)=T(|k|,|k'|;E,Q)$. The two integration domains then combine into the full real axis.

In either term, the intermediate three-magnon energy is
\begin{equation}
u|q_\eta|^z+u|q_\eta'|^z+u|q_\eta+q_\eta'|^z,
\end{equation}
which gives the free propagation kernel
\begin{equation}
K(q_\eta,q_\eta';E)
=\frac{1}{E-u|q_\eta|^z-u|q_\eta'|^z-u|q_\eta+q_\eta'|^z}.
\end{equation}
Combining these expressions yields
\begin{equation}
\begin{aligned}
\phi(q_\xi,q_\eta)
={}&\int_{-\infty}^{\infty}\frac{dq_\eta'}{2\pi}
T\left(q_\xi,q_\eta'+\frac{q_\eta}{2};
E-u|q_\eta|^z,-q_\eta\right)\\
&\times K(q_\eta,q_\eta';E)
\phi\left(-q_\eta-\frac{q_\eta'}{2},q_\eta'\right),
\end{aligned}
\label{eq:app_STM}
\end{equation}
in agreement with Eq.~\eqref{eq:STM}.

\bibliography{main.bbl}

\end{document}